\documentclass[twoside,twocolumn,9pt]{article}
\usepackage{extsizes}
\usepackage[super,sort&compress,comma]{natbib} 
\usepackage[version=3]{mhchem}
\usepackage[left=1.5cm, right=1.5cm, top=1.785cm, bottom=2.0cm]{geometry}
\usepackage{balance}
\usepackage{mathptmx}
\usepackage{sectsty}
\usepackage{graphicx} 
\usepackage{lastpage}
\usepackage{multicol}
\usepackage{svg}
\usepackage[format=plain,justification=justified,singlelinecheck=false,font={stretch=1.125,small,sf},labelfont=bf,labelsep=space]{caption}
\usepackage{float}
\usepackage{fancyhdr}
\usepackage{fnpos}
\usepackage[english]{babel}
\addto{\captionsenglish}{%
  
}
\usepackage{array}
\usepackage{droidsans}
\usepackage{charter}
\usepackage[T1]{fontenc}
\usepackage{setspace}
\usepackage[compact]{titlesec}
\usepackage[colorlinks,allcolors=blue]{hyperref}
\usepackage{float}
\usepackage{tikz}

\usepackage{epstopdf}

\definecolor{cream}{RGB}{222,217,201}

\begin{document}

\pagestyle{fancy}
\thispagestyle{plain}
\fancypagestyle{plain}{
\renewcommand{\headrulewidth}{0pt}
}

\makeFNbottom
\makeatletter
\renewcommand\LARGE{\@setfontsize\LARGE{15pt}{17}}
\renewcommand\Large{\@setfontsize\Large{12pt}{14}}
\renewcommand\large{\@setfontsize\large{10pt}{12}}
\renewcommand\footnotesize{\@setfontsize\footnotesize{7pt}{10}}
\makeatother

\renewcommand{\thefootnote}{\fnsymbol{footnote}}
\renewcommand\footnoterule{\vspace*{1pt}%
\color{cream}\hrule width 3.5in height 0.4pt \color{black}\vspace*{5pt}} 
\setcounter{secnumdepth}{5}

\makeatletter 
\renewcommand\@biblabel[1]{#1}            
\renewcommand\@makefntext[1]%
{\noindent\makebox[0pt][r]{\@thefnmark\,}#1}
\makeatother 
\renewcommand{\figurename}{\small{Fig.}~}
\sectionfont{\sffamily\Large}
\subsectionfont{\normalsize}
\subsubsectionfont{\bf}
\setstretch{1.125} 
\setlength{\skip\footins}{0.8cm}
\setlength{\footnotesep}{0.25cm}
\setlength{\jot}{10pt}
\titlespacing*{\section}{0pt}{4pt}{4pt}
\titlespacing*{\subsection}{0pt}{15pt}{1pt}

\fancyfoot{}
\fancyfoot[LO,RE]{\vspace{-7.1pt}\includegraphics[height=9pt]{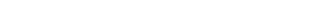}}
\fancyfoot[CO]{\vspace{-7.1pt}\hspace{13.2cm}\includegraphics{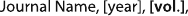}}
\fancyfoot[CE]{\vspace{-7.2pt}\hspace{-14.2cm}\includegraphics{head_foot/RF}}
\fancyfoot[RO]{\footnotesize{\sffamily{1--\pageref{LastPage} ~\textbar  \hspace{2pt}\thepage}}}
\fancyfoot[LE]{\footnotesize{\sffamily{\thepage~\textbar\hspace{3.45cm} 1--\pageref{LastPage}}}}
\fancyhead{}
\renewcommand{\headrulewidth}{0pt} 
\renewcommand{\footrulewidth}{0pt}
\setlength{\arrayrulewidth}{1pt}
\setlength{\columnsep}{6.5mm}
\setlength\bibsep{1pt}

\makeatletter 
\newlength{\figrulesep} 
\setlength{\figrulesep}{0.5\textfloatsep} 

\newcommand{\topfigrule}{\vspace*{-1pt}%
\noindent{\color{cream}\rule[-\figrulesep]{\columnwidth}{1.5pt}} }

\newcommand{\botfigrule}{\vspace*{-2pt}%
\noindent{\color{cream}\rule[\figrulesep]{\columnwidth}{1.5pt}} }

\newcommand{\dblfigrule}{\vspace*{-1pt}%
\noindent{\color{cream}\rule[-\figrulesep]{\textwidth}{1.5pt}} }

\makeatother

\twocolumn[
  \begin{@twocolumnfalse}
{\includegraphics[height=30pt]{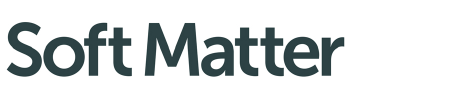}\hfill\raisebox{0pt}[0pt][0pt]{\includegraphics[height=55pt]{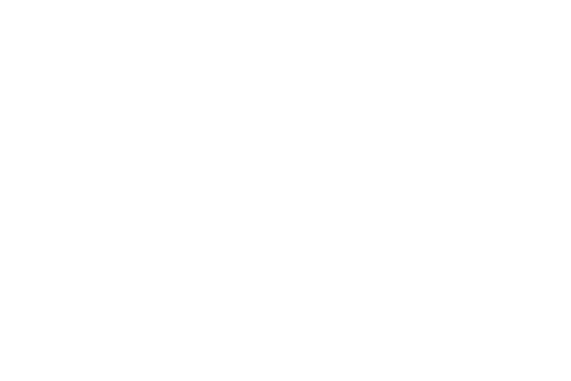}}\\[1ex]
\includegraphics[width=18.5cm]{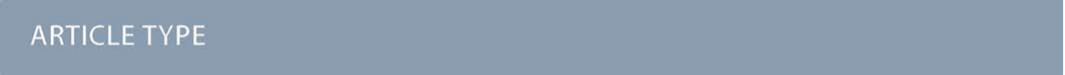}}\par
\vspace{1em}
\sffamily
\begin{tabular}{m{4.5cm} p{13.5cm} }

\includegraphics{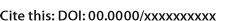} & \noindent\LARGE{\textbf{Shear zones in granular mixtures of hard and soft particles with high and low friction}}\\
\vspace{0.3cm} & \vspace{0.3cm} \\

 & \noindent\large{Aditya Pratap Singh, Vasileios Angelidakis, Thorsten P\"oschel, and Sudeshna Roy\textit{$^{\dag}$}} \\

\includegraphics{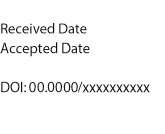} & \noindent\normalsize{Granular materials show inhomogeneous flows characterized by strain localization. When strain is localized in a sheared granular material, rigid regions of a nearly undeformed state are separated by shear bands, where the material yields and flows. The characteristics of the shear bands are determined by the geometry of the system, the micromechanical material properties, and the kinematics at the particle level. For a split-bottom shear cell, recent experimental work has shown that mixtures of hard, frictional and soft, nearly frictionless particles exhibit wider shear zones than samples with only one of the two components. To explain this finding, we investigate the shear zone properties and the stress response of granular mixtures using discrete element simulations. We show that both interparticle friction and elastic modulus determine the shear-band properties of granular mixtures of various mixing ratios, but their stress response depends strongly on the interparticle friction. Our study provides a fundamental understanding of the micromechanics of shear band formation in granular mixtures.} \\
\end{tabular}

 \end{@twocolumnfalse} \vspace{0.6cm}

  ]

\renewcommand*\rmdefault{bch}\normalfont\upshape
\rmfamily
\section*{}
\vspace{-1cm}


\footnotetext{\textit{Institute for Multiscale Simulation, Friedrich-Alexander-Universit\"at Erlangen-N\"urnberg, Cauerstrasse 3, 91058 Erlangen, Germany}}

\footnotetext{\dag~sudeshna.roy@fau.de}


\section{Introduction}
Strain localization is a characteristic effect of sheared granular matter, which is not observed in the flow of other materials, such as liquids. Once strain localizes within a narrow shear zone (also referred to as the shear band), the stress, strain, and strain rate of the material assume high values within that zone. On the contrary, outside of it, the material remains nearly undeformed. The knowledge of the properties of shear zones and the factors that influence them is crucial for understanding the intrinsic properties of granular materials under shear. 

In many systems where particles come into contact with boundaries, narrow shear zones appear at the interface because of particle-boundary interactions. These shear zones are triggered by boundary effects, which hinder the accurate characterization of bulk material properties based on the shear zone characteristics. In our work, we present a unique setup called a split-bottom shear cell, which is a modified version of a Couette cell. This setup involves splitting the cell's bottom at a specific radius ($R_s$), resulting in two parts consisting of the inner disk and outer cylinder components (see \autoref{fig:schematic}). After loading the cell with grains to a certain height ($H$), the inner disk is set to rotational motion.
This design allows granular materials to be sheared to develop wide shear 
zones inside the bulk of the materials, far away from the rigid boundaries, which allows for a comprehensive characterization of their yielding properties and flow behavior at a macromechanical level. This is quite different from the usual narrow shear bands that we observe close to the boundary, which are formed due to particle-boundary interactions. This shear cell setup has been used in experiments \cite{wortel2015heaping} and simulations \cite{fischer2016heaping} to explore the bulk properties of granular systems.

The universal geometrical properties of shear bands in the split-bottom shear cell have been studied extensively in the literature \cite{fenistein2003wide,fenistein2004universal,dijksman2010granular}. 
The absolute value of the macroscopic normalized angular velocity of the granular material was measured as a function of the radial position, and it has been found to follow closely the shape of an error function, 
which is characterized by two parameters: the width $W$ and the center position $R_c$ of the shear zone \cite{dijksman2010granular,fenistein2004universal}. The width of the shear band increases with increasing height within the bulk of the material, while the distance of its central position $R_c$ from the center of rotation decreases (that is, the shear band moves inward with increasing height). The width of the shear band is affected by the size, shape, and interaction properties of the particles, such as friction between the particles, cohesion, and the shear rate \cite{henann2013predictive,singh2014effect,luding2007effect,faroux2022granular}. The position of the shear band, $R_c$, at the surface of the granular material in the shear cell, 
depends on two geometrical parameters, the filling height $H$ and split radius $R_s$ \cite{unger2004shear}. The bulk profile of the shear band as a function of height, $R_c(z)$, can be linked to a variational problem. The solution of the function $R_c(z)$ that minimizes the integral representing the dissipation rate, is related to the minimal mechanical torque and thus depends on the material properties. Therefore, the shape of the shear band within the bulk is not only influenced by geometric factors such as the split position ($R_s$) and the filling height ($H$) but is also closely connected to the material properties, e.g., the shear stress, and the packing fraction of the material, as highlighted in previous studies \cite{unger2004shear, dijksman2010granular}.

Interparticle friction and particle softness play a key role in influencing material bulk flows, which have been widely studied. Wang et al. \cite{wang2022characterization} experimentally characterized the shear zones of binary mixtures made of hard, frictional and soft, slippery particles. Their study identified distinct shear band characteristics specific to the mixture, distinguishing them from those in pure-species systems. Luding \cite{luding2007effect} explored the impact of interparticle friction on shear zones in split-bottom shear cells through numerical simulations. Increasing friction resulted in narrower shear bands that shifted towards the center. Ashour et al. \cite{ashour2017silo} found that particle softness adds qualitatively new features to the dynamics and the packing characteristics of silo flows. Soft, deformable particles, such as hydrogel particles, can adapt to local stress conditions, leading to alterations in pressure profiles and flow fields. Hong et al. \cite{hong2017clogging} highlighted the importance of particle softness in hopper flows, where gravitational forces must be less than repulsive forces from particle stiffness to induce clogging. Harth et al. \cite{harth2020intermittent} observed longer time scales for internal rearrangement in soft hydrogel particles during silo orifice flow compared to hard grain systems, where shorter time scales have been reported by Unac et al. \cite{unac2012experimental}. This was attributed to mass flow rate differences, roundness, and increased viscous damping of hydrogel grains. The rate-dependent frictional regimes observed in hydrogel particle suspensions have been attributed to their high deformability and low friction \cite{rudge2020uncovering}. Wang et al. \cite{wang2021silo} observed experimentally that, unlike granular materials with hard particles, soft slippery particles in a silo flow do not follow Beverloo's \cite{beverloo1961flow} law, according to which the flow rate is independent of the overburden pressure. Adding just $5\%$ hard particles to soft silo flows restored this behavior and significantly altered system flow properties. Based on the micromechanical observations, Saitoh et al. \cite{saitoh2015master} highlighted that the evolution of the system behavior in soft particle packings extends beyond just particle contacts. Inter-particle gaps between nearest neighbors are also crucial and affect the stochastic evolution of the probability distribution function of forces. Recently, Luding et al. \cite{luding2022understanding} studied the contact and force networks of frictionless soft granular materials, revealing that the evolution of the pressure in the system was correlated with the existence of loops formed in the force network during particle rearrangements. G\"otz et al. \cite{gotz2022soft} observed that soft particles with larger deformations and increased contact areas result in larger frictional forces and improved holding capacity of a granular gripper. In summary, both particle friction and softness are crucial factors influencing bulk flow behavior, and the intricate interplay between these properties and their combined effects on bulk behavior remains a subject of ongoing investigation.

The findings of Wang et al. \cite{wang2022characterization} established that mixtures of hard and soft particles exhibit wider shear zones than assemblies made of only one of the individual materials (only hard or only soft particles). However, the underlying reason for this behavior remains unclear. Our objective here is to gain insight into the combined effect of microscopic parameters, namely particle stiffness and inter-particle friction, on the macroscopic shear zone characteristics for ensembles of soft, slippery particles and hard, frictional particles sheared in a split-bottom shear cell.
\section{Numerical setup and materials}\label{sec:setup}
We use an open-source Discrete Element Method (DEM) code MercuryDPM \cite{weinhart2020fast} to simulate sheared systems of hard and soft particles. The numerical setup used for this study is a split-bottom shear cell,
as shown in \autoref{fig:schematic}, where the inner bottom plate rotates around a central axis at a constant angular velocity, $\omega_o = 0.52$ rad/s, while the outer boundary of the cell remains stationary. The granular material located in the outer parts remains static, while the material near the center of the cell rotates with the bottom disk. The shear in this system is localized in an intermediate region, defining a wide shear band. The position and the width of the shear zone depend on the ratio $H/R_s$, where $H$ is the filling height and $R_s$ is the split radius of the rotating bottom plate \cite{sakaie2008mr,fenistein2004universal}. For small and moderate filling height $(H < 0.6 R_s)$, the shear zone starts from the bottom split position, evolves through the entire height of the granular layer, and ends up at the surface \cite{fenistein2003wide}, as shown in \autoref{fig:schematic}, while the inner part of the granular material co-rotates synchronously with the bottom plate. In our numerical setup, the filling height is $H = 20$ mm and  $H/R_s = 0.42$, which is considered a moderate filling height. {
This numerical setup is similar to that employed in the experimental work by Wang et al. \cite{wang2022characterization}, with the difference that the experimental configuration incorporates an additional small rod in the center of the cell, where the shear load was imposed.

\begin{figure}
    \begin{center}
        \includegraphics[width=0.9\columnwidth]{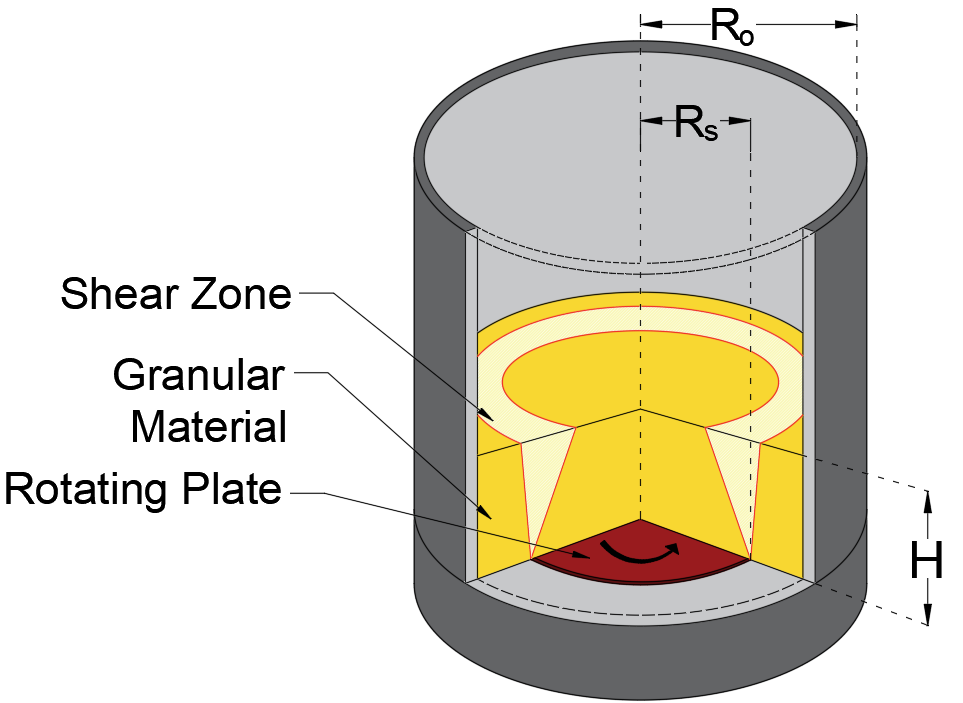}
    \end{center}
    \caption{Schematic of a split bottom shear cell of radius $R_o$ with granular material at filling height $H$, illustrating the main system components. The bottom plate (dark-red color) rotates around the vertical axis at constant rotation frequency $\omega_o$, while the outer wall (marked in grey) stays stationary. The differential motion triggers the formation of a shear band (marked in orange color) starting from the split radius $R_s$, which widens with increasing height. In contrast, the material outside the shear band (marked in yellow) stays undeformed.
    }
    \label{fig:schematic}
\end{figure}

\subsection{Material parameters}
We consider a mixture of soft low-frictional particles and hard frictional particles. In this Section, we describe the generation of binary granular mixtures of such particles. We take the properties of hydrogel spheres as the soft, low-frictional material component, and mustard seeds represent hard, frictional particles. A reference sample of pure soft hydrogel particles consists of $N^0_A = 26795$ polydisperse particles of species A with mean diameter, $d_A = 2.5$ mm. Starting from this baseline sample volume, we create mixtures by substituting $N^0_A - N^T_A$ particles of species A with particles of species B of diameter $d_B = 3$ mm, such that the same volume of material A is replaced by B, i.e. $V_B = (N^0_A - N^T_A)(\pi/6)d_A^3 = N^T_B(\pi/6)d_B^3$. For both species A and B, we use particles of homogeneous size distribution of mean particle diameter $d_A$ and $d_B$, respectively, with polydispersity of $\pm 8\%$. To characterize the mixtures with different ratios of species B, we define the mixing ratio
\begin{equation}
    X = \frac{N^T_B}{N^T_A + N^T_B}
\end{equation}
We analyze systems with $X \in \{0, 0.05, 0.25, 0.50, 0.75, 0.85, 1.0\}$. \autoref{tab:NParticles} presents the number of particles of the two species in each sample. \autoref{fig:topview} shows some examples of different mixtures. 
 
The properties of soft particles of species A correspond to that of hydrogel spheres with friction coefficient $\mu_A = 0.005$ and elastic modulus $E_A = 50$ kPa. The frictional properties of hard particles correspond to mustard seeds with friction coefficient $\mu_B = 0.20$ and the elastic modulus $E_B = 5000$ kPa. All relevant material and interaction properties of the two species are given in \autoref{tab:matprop}.

\begin{table} 
\small
  \caption{\ Number of particles of the species in each sample}
    \label{tab:NParticles}
  \begin{tabular*}{0.48\textwidth}{@{\extracolsep{\fill}}llllllll}
    \hline
    $X$ & $0$ & $0.05$ & $0.25$ & $0.50$ & $0.75$ & $0.85$ & $1$ \\
    Species A & $26975$ & $24675$ & $17104$ & $9757$ & $4244$ & $2497$ & $0$ \\
    Species B & $0$ & $1361$ & $5695$ & $9917$ & $13160$ & $14133$ & $15465$ \\
    \hline
  \end{tabular*}
\end{table}
\begin{table}[htb!] 
\small
  \caption{\ Microscopic material parameters for the model}
    \label{tab:matprop}
  \begin{tabular*}{0.48\textwidth}{@{\extracolsep{\fill}}lll}
    \hline
    Properties & Species A & Species B \\
    \hline
    Mean particle diameter & $2.5$ mm & $3$ mm \\
    Density & $1017$ kg/m$^{-3}$ & $1017$ kg/m$^{-3}$ \\
    Elastic modulus & $50$ kPa & $5000$ kPa \\
    Poisson's ratio & $0.40$ & $0.40$ \\
    Friction coefficient & $0.005$ & $0.20$ \\ 
    \hline
  \end{tabular*}
\end{table}

\begin{figure*}
    \centering
    {\includegraphics[width=\textwidth, trim={3cm 16cm 3cm 10.5cm},clip]{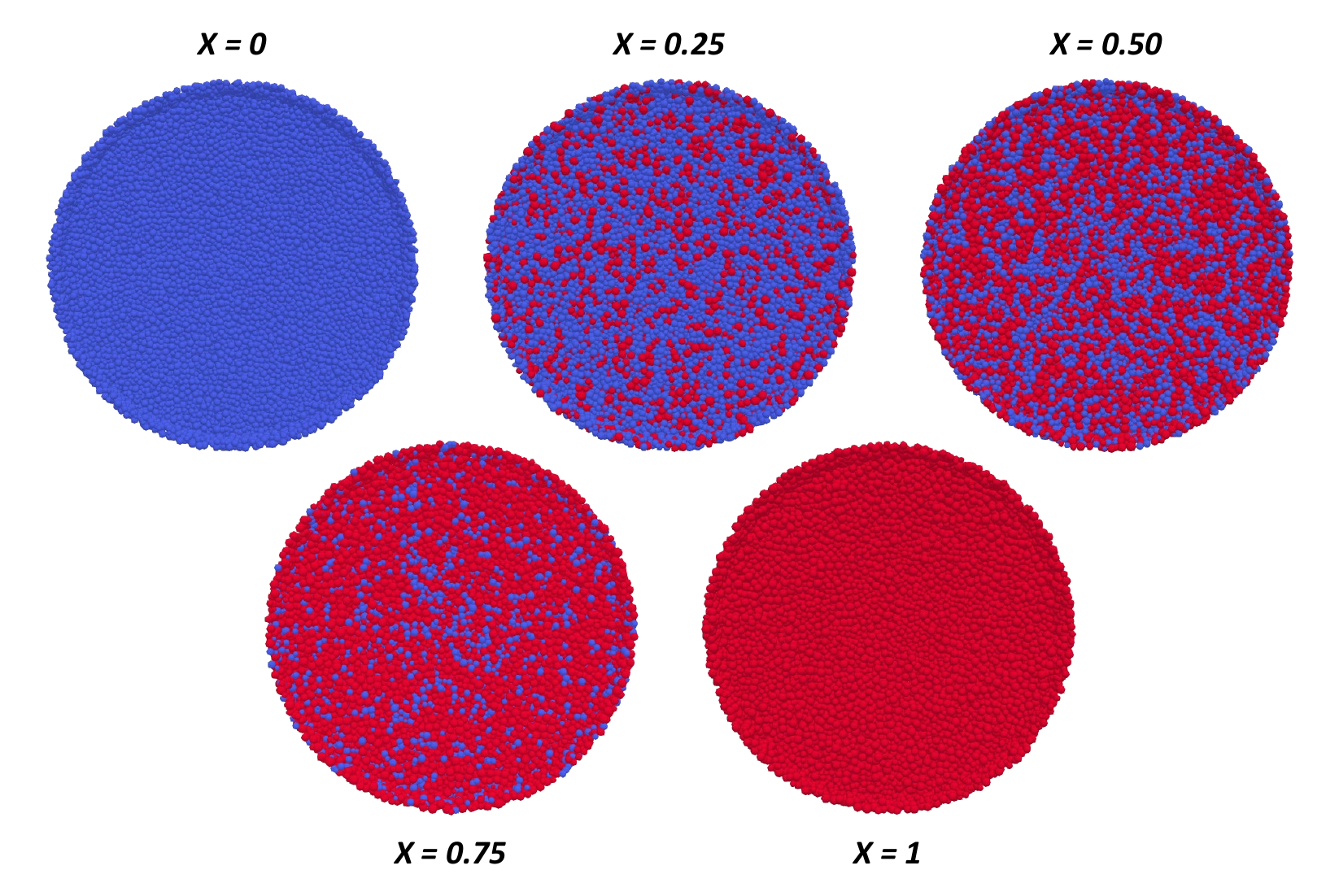}}
    \caption{Top view of the shear cell for granular mixtures of different percentages of hard particles $X$, where the blue and red particles represent species A (soft) and B (hard), respectively.
    }
    \label{fig:topview}
\end{figure*}

\subsection{Contact model}
In the soft-particle DEM, the evolution of particle positions and velocities is calculated by integrating Newton’s equation of motion. When two particles $i$ and $j$ interact, they can slightly overlap. Only viscoelastic interactions are considered, i.e., after separating the contacting particles, they recover their initial shape. The compression of contacting particles $\xi$ is written as:
 
\begin{equation}
    \xi = R_i + R_j - |\vec{r_i} - \vec{r_j}|
\end{equation}
where $R_i$, $R_j$ are the particle radii and $\vec{r_i}$, $\vec{r_j}$ are the particle positions. The normal force is calculated as in \cite{brilliantov1996model}, where an elastic Hertzian normal force is considered in conjunction with a viscous damper of velocity-dependent coefficient of restitution.

\begin{equation}
    \vec{F_n} = \min\left(0, -\rho \xi^{3/2} - \frac{3}{2}A_n\rho\sqrt{\xi}\dot{\xi}\right) \vec{e_n}
\end{equation}
with

\begin{equation}
    \rho = \frac{4}{3} \, E^{*} \, \sqrt{R^*} 
    \label{Eq:Hertz_rho}
\end{equation}

where $E^*$ is the effective elastic modulus, $R^*=R_i R_j/\left(R_i+R_j\right)$ is the effective radius, 
$\vec{e}_n = (\vec{r}_i-\vec{r}_j)/\vert\vec{r}_i-\vec{r}_j\vert$ is the normal unit vector, and $A_n = 4.1\times 10^{-6}$ s is the normal dissipative parameter, calculated as in \cite{muller2011collision}, considering a coefficient of restitution of $0.70$ for the tangential translational velocity of the rotating bottom plate at the split radius. 
The effective elastic modulus $E^*$ is given by
 \begin{equation}
     E^* = {\bigg(\frac{1-\nu_i^2}{E_i} + \frac{1-\nu_j^2}{E_j}\bigg)}^{-1}
\end{equation}
where $E_{i}$, $E_{j}$ and $\nu_{i}$, $\nu_{j}$ are the elastic moduli and Poisson ratios of particles $i$, $j$, respectively.

We model the tangential viscoelastic forces considering a path-dependent calculation, following the no-slip solution of Mindlin \cite{mindlin1949compliance} for the elastic part and Parteli and P\"oschel \cite{parteli2016particle} for the tangential dissipative parameter $A_t \approx 2 A_n E^*$, which are capped by the static friction force between two particles. The tangential force is thereby given by
\begin{equation}
    \vec{F_t} = -\min \left[ \mu|\vec{F_n}|,  \int_\mathrm{path}^{} 8 G^{*}\sqrt{R^* \xi} \,ds 
    + A_t  \sqrt{R^* \xi} v_t \right] \vec{e_t}
\end{equation}
where $\mu$ is the friction coefficient, $\vec{e_t}$ is the unit vector in the tangential contact direction and $G^*$ the effective shear modulus, given by
\begin{equation}
    G^* = {\bigg(\frac{2-\nu_i}{G_i} + \frac{2-\nu_j}{G_j}\bigg)}^{-1}    
\end{equation}
The shear modulus $G_i = \frac{E_i}{2(1+\nu_i)}$ is calculated assuming isotropic material. For the interaction between different species, the effective friction coefficient is considered as $\mu^* = 2{\bigg(\frac{1}{\mu_i}+\frac{1}{\mu_j}\bigg)}^{-1}$. It is typical for many DEM formulations to consider the minimum coefficient of friction when two particles of different materials come into contact. Since the harmonic average of two values favors the smaller one, considering this calculation of the friction coefficient for contacts between high and low friction coefficients gives values close to considering the minimum value. This assumption is chosen as a milder approach to simply assuming the minimum friction coefficient of the two, and more experimental evidence is required to understand the friction between hard and soft particles. Considering the harmonic average here, we have three scenarios of possible contacts between frictional-frictional, slippery-slippery and frictional-slippery (with friction very close to slippery-slippery) particles.
\section{Segregation effects}
\begin{figure*}
    \centering
    \begin{minipage}{0.5\textwidth}
        \begin{picture}(100,170)
        \put(0,0){\includegraphics[width=1.0\textwidth]{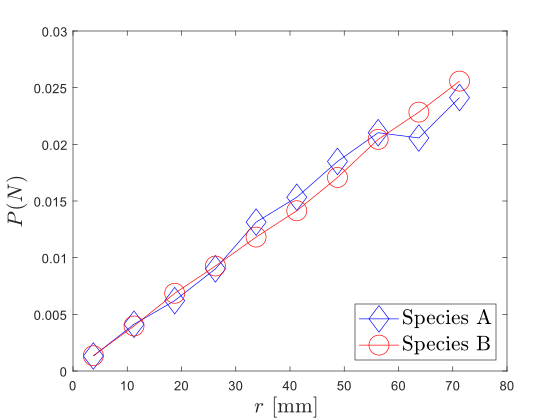}}
        \put(60,150){(a)}
        \end{picture}
    \end{minipage}\hfill
    \begin{minipage}{0.5\textwidth}
        \begin{picture}(100,170)
        \put(0,0){\includegraphics[width=1.0\textwidth]{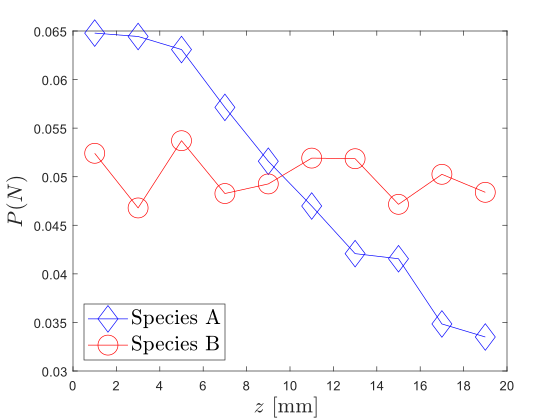}}
        \put(180,150){(b)}
        \end{picture}
    \end{minipage}\hfill
    \caption{Probability of particle number as a function of (a) radial distance and (b) distance from the bottom surface for species A and B after $200$ s of simulation time.}\label{fig:segregation}
\end{figure*}
When granular mixtures with particles differing in size and/or density are subjected to external forces such as shaking, stirring, or shearing, they often segregate to form complex patterns. These patterns of segregation can lead to different and inhomogeneous macroscopic behavior. In many applications where a homogeneous packing is sought, this is undesirable, and significant effort is often invested in attempts to avoid or control segregation during processing and handling of granular materials. Shear-induced percolation segregation that occurs due to \textit{kinetic sieving} and \textit{squeeze expulsion} is the dominant mechanism in dense granular flows. In this study, we study a bi-disperse mixture of two species that differ slightly in terms of particle size. Therefore, in this section, we examine the extent to which particles segregate in this system of hard frictional and soft slippery particles, sheared in the split-bottom shear cell.

\begin{figure*}
    \centering
    \begin{minipage}{0.5\textwidth}
        \begin{picture}(100,170)
            \put(0,0){\includegraphics[width=1.0\textwidth]{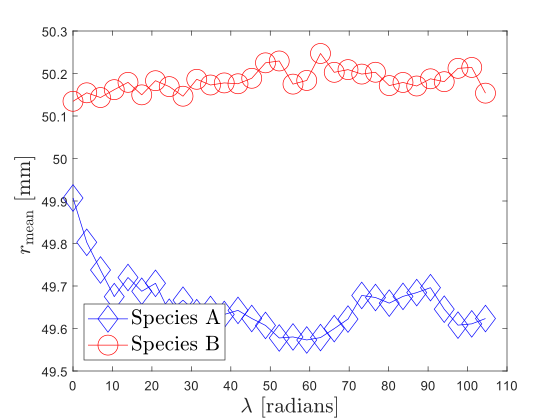}}
            \put(185,80){(a)}
        \end{picture}
    \end{minipage}\hfill
    \begin{minipage}{0.5\textwidth}
        \begin{picture}(100,170)
            \put(0,0){\includegraphics[width=1.0\textwidth]{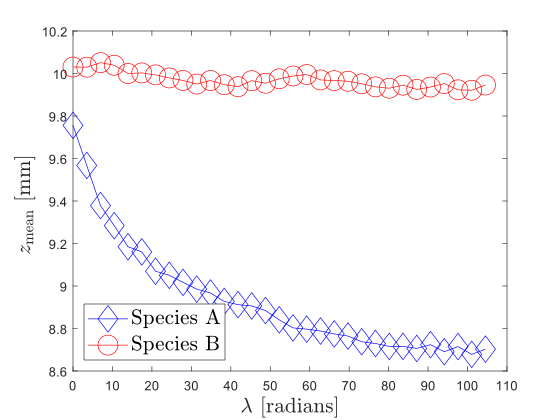}}
            \put(185,60){(b)}
        \end{picture}
    \end{minipage}\hfill
    \caption{Average (a) radial distance and (b) distance from the bottom surface for the species A and B as a function of shear displacement $\lambda = \omega t_\mathrm{rot}$.}\label{fig:segregationTime}
\end{figure*}
First, we analyze the probability distribution of the particles of each species in the radial $r$ and vertical $z$ directions for a given snapshot after $200$ s of simulation time, as shown in \autoref{fig:segregation}(a) and (b), respectively. The distribution increases in the radial direction for both species due to the increased measuring volume (the probability is calculated using co-centric cylinders of increasing radius, aligned with the center of the cell). Note that the distribution of the two species overlaps with each other in the radial $r$ direction, indicating that there is no segregation effect of the two species horizontally. The distribution in the vertical $z$ direction for species B is nearly constant. However, the distribution of species A decreases in vertical $z$ direction from $0.065$ to $0.035$, indicating a higher probability of species A near the bottom of the shear cell than near the free surface. This is a signature of the segregation of species A in the vertical direction due to \textit{kinetic sieving}, and is attributed to the smaller particle size compared to species B. To further analyze this effect, \autoref{fig:segregationTime}(a) and (b) illustrate the mean radial positions $r$ and perpendicular positions $z$ of the two species as a function of shear displacement $\lambda$.
Here, it is to be noted that the mean position of Species A is about $0.4$ mm less in the $r$ direction and $1.2$\ mm less in the $z$ direction than that of Species B. However, the differences are negligibly small and of less than one particle diameter.

\section{Micro-macro transition}
The current work aims to derive appropriate macroscopic fields commensurate with the shear band features and macromechanical stress analysis, e.g., strain rate, velocity profile, and stress, based on the given micromechanical properties. The flow that rapidly relaxed to a steady state is purely azimuthal and is proportional to the driving rate of the shear cell, $\omega_o$. The averaging is thus performed in toroidal volumes over many snapshots of time, leading to generic continuum fields $Q(r,z)$ as a function of the radial and vertical positions. We choose the representative elementary volume according to the \cite{latzel2000macroscopic} as equal to the minimum particle diameter $d_p = 1.37$ mm.

Shearing of a granular system leads to volume dilation and build-up of shear stress and anisotropy in fabric. Since we are interested in the steady-state flow profile, we probe for the minimum strain required to reach the steady-state flow regime. We define the shear displacement $\lambda = \omega t_\mathrm{rot}$ through which the system rotates (where $t_\mathrm{rot}$ is the simulation time) as the shear control parameter. As reported by Singh et al. \cite{singh2015role}, both the global quantities of kinetic energy and average number of contacts reach a steady state very fast ($\lambda \sim 5$). We analyze the relaxation of local quantities, specifically the local velocity profiles, and conclude that $\lambda \sim 20-30$ is the amount of shear displacement required to form a stable shear band, which is in agreement with Ries et al. \cite{ries2007shear}. The data depicting the initial, transient evolution of local shear band properties is not presented here, as our focus is solely on the steady-state characteristics of the shear band. Consequently, we perform the local averaging over almost $600$ snapshots distributed over $\lambda \sim 80$ to $100$.
 
The consistency of the local averaged quantities also depends on the accumulated local shear strain during the averaging time. We concentrate our interest in the region where the system can be considered to reach a critical state. The critical state is a unique steady state reached after extensive shear, where the material deforms with applied strain without any further change in normal stress, shear stress, and volume fraction. This is a state where the system forgets its loading, packing, or other sample-preparation history \cite{wood1990soil}. In accordance with previous experimental and numerical results in the same setup \cite{luding2008constitutive,szabo2014evolution}, the criterion of large strain rate identifies the shear band region, e.g., higher than a critical strain rate of $\dot\gamma_c = 0.08$ s$^{-1}$. Here, the shear band center region is defined by strain rates higher than 80\% of the maximum strain rate for different heights in the shear cell. Therefore, the critical strain rate here is a function of height in the shear cell and is denoted as $\dot\gamma_c (z) = 0.8{ }\dot\gamma_\mathrm{max}(z)$.

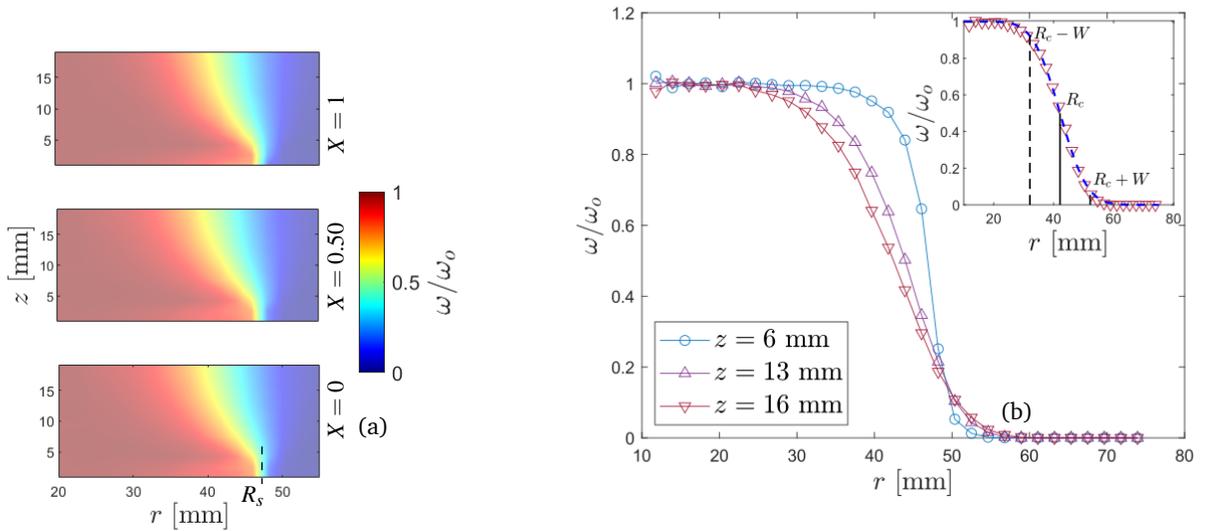
\begin{figure*}
    \centering
    \begin{minipage}{0.5\textwidth}
        \begin{tikzpicture}
            \node (0,0) {\includegraphics[width=\textwidth]{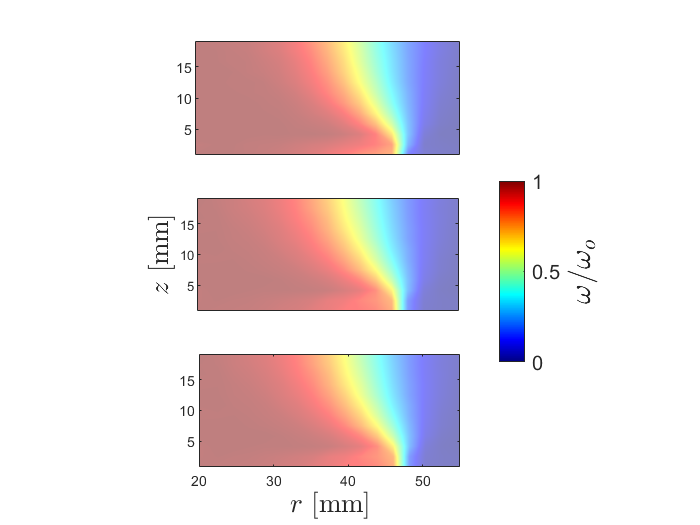}}; 
            \draw[dashed, line width=0.5pt] (0.7,-2.3) -- ++(0,-0.5cm);
            \node[anchor=south east] at (0.85,-3.2) {$R_s$};
            \node[anchor=south east] at (1.9,-2.3) {\rotatebox{90}{$X = 0$}};
            \node[anchor=south east] at (1.9,-0.6) {\rotatebox{90}{$X = 0.50$}};
            \node[anchor=south east] at (1.9,1.5) {\rotatebox{90}{$X = 1$}};
            \node[anchor=south east] at (2.5,-2.3) {(a)};
        \end{tikzpicture}
    \end{minipage}\hfill
    \begin{minipage}{0.5\textwidth}
        \begin{picture}(100,170)
        \put(0,0){\includegraphics[width=1.0\textwidth]{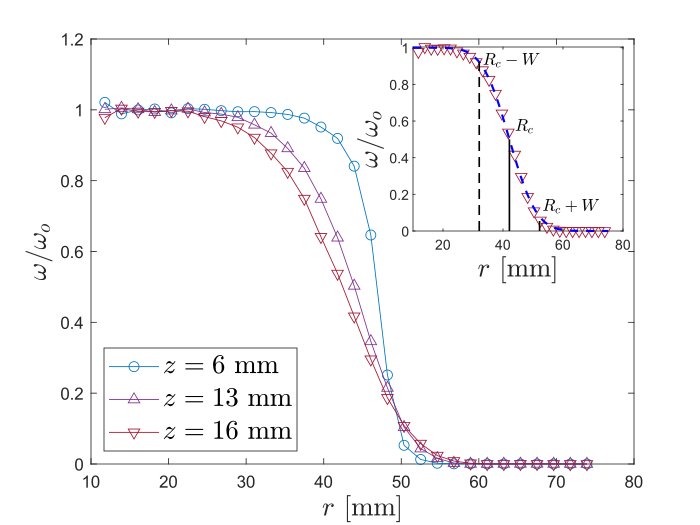}}
        \put(170,30){(b)}
    \end{picture}
    \end{minipage}\hfill         
    \caption{(a) Shear band profiles for mixtures of soft and hard particles in different mixing ratios. The color scale illustrates the normalized ratio of macroscopic angular velocity $\omega(r,z)$ scaled by the rotation velocity of the bottom disk $\omega_o$. $R_s$ marks the split radius, which serves as the origin of the shear band in space. (b) Graphical representation of macroscopic angular velocity $\omega(r)$ normalized by the bottom disk angular velocity $\omega_o$, as a function of radial position $r$ for a fraction of hard particles $X = 0.50$ at different heights. The inset shows the scaled angular velocity $\omega(r)/\omega_o$ vs. radial position $r$ fitted by \autoref{eq:SBSfit}, also showing the features of the shear band.}\label{fig:scaledOmega}            
\end{figure*}
\section{Flow profile of granular mixtures}
Once the flow of the mixtures relaxes to steady state shearing, the azimuthal velocity is proportional to the driving rate $\omega_o$. {
\autoref{fig:scaledOmega}(a) shows the scaled angular velocity profile $\omega(r,z)/\omega_o$. Within the limits of assumptions in our simulations, we do not see a clear difference in the flow profiles of the three cases of pure particles and mixtures for $X = 0, 0.50$, and $1$. However, we observe that the shear zone of the mixture $(X=0.5)$ is slightly more diffused compared to the pure case simulations involving only soft $(X=0)$ or only hard particles $(X=1)$. This indicates a wider shear band for the mixture, although it is not clearly concluded from this figure. A closer and more precise introspection on the shear band properties is therefore essential to conclude the shear zone properties of the mixture. On this ground, we analyze the dimensionless ratio of the measured angular velocity $\omega/\omega_o$ as a function of the radial coordinate $r$. The dimensionless velocity profile in a split bottom shear cell is well approximated by an error function \cite{dijksman2010granular,unger2004shear}}

\begin{equation}\label{eq:SBSfit}
    \frac{\omega(r)}{\omega_o} = 0.5 - 0.5\mathrm{erf}\bigg{(}\frac{r-R_c}{W}\bigg{)}
\end{equation}
where $R_c$ and $W$ are the position and width of the shear band developed above the split at $R_s$, respectively. Both the position and the width of the shear band depend on the height $z$ in the system. Therefore, we extract the position and the width of the shear band at layers of varying height along the filling height in the shear cell by fitting the scaled $\omega(r)$ against the radial position $r$ data. With the goal to extract quantitative data for the shear band features, \autoref{fig:scaledOmega}(b) shows values of non-dimensional angular velocity $\omega(r)/\omega_o$ at different heights in the bulk of granular materials against the radial distance.

A quantitative study of the velocity gradient leads to the strain rate {$\dot\gamma$, which is calculated from the shear-free sheet strain tensor components $d_1$ and $d_2$ \cite{depken2006continuum}. The shear intensity in the shear plane is thus expressed in terms of velocity gradients ${\partial v_\theta}/{\partial r}$ and ${\partial v_\theta}/{\partial z}$ as}:

\begin{equation}\label{eq:strainrate}
    \dot\gamma = \sqrt{d_1^2+d_2^2} = \frac{1}{2}\sqrt{{\bigg(\frac{\partial v_\theta}{\partial r} - \frac{v_\theta}{r}\bigg)}^2 + {\bigg(\frac{\partial v_\theta}{\partial z}\bigg)}^2}
    \end{equation}
\autoref{fig:shearbandwidth}(a) shows the strain rate $\dot\gamma(r,z)$ evaluated from \autoref{eq:strainrate} and plotted as a function of radial and vertical position for the sample with $X = 0.50$. The dashed lines overlaid on top of the contour graph indicate the center $R_c$ and the edges $R_c+W$ and $R_c-W$ of the shear band, indicated in the figure, as obtained from the fit function in \autoref{eq:SBSfit}.

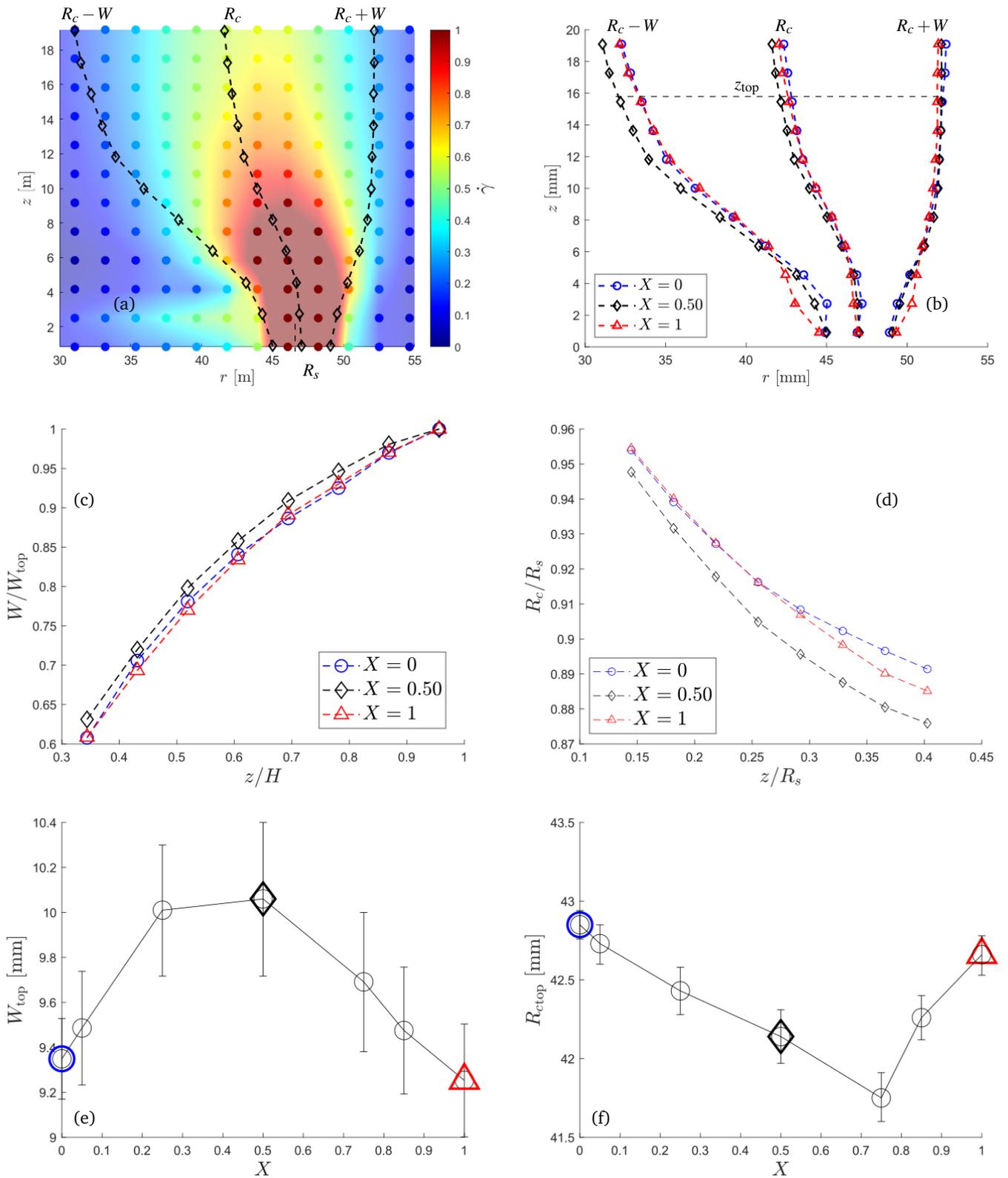
\begin{figure*}
    \centering
    \begin{minipage}{0.5\textwidth}
        \begin{tikzpicture}
            \node (-1,0) {\includegraphics[width=\textwidth]{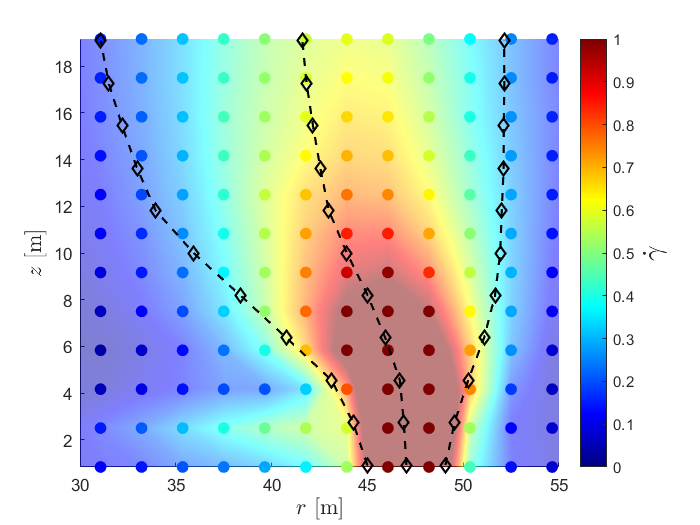}}; 
            \draw[dashed, line width=0.5pt] (0.64,-2.3) -- ++(0,-0.5cm);
            \node[anchor=south east] at (1.2,-3.4) {$R_s$};
            \node[anchor=south east] at (-0.2,3) {$R_c$};
            \node[anchor=south east] at (2.4,3) {$R_c+W$};
            \node[anchor=south east] at (-2.5,3) {$R_c-W$};
            \node[anchor=south east] at (-2.1,-2.2) {(a)};
        \end{tikzpicture}
    \end{minipage}\hfill
    \begin{minipage}{0.5\textwidth}
        \begin{tikzpicture}
       \node (-1,0) {\includegraphics[width=\textwidth]{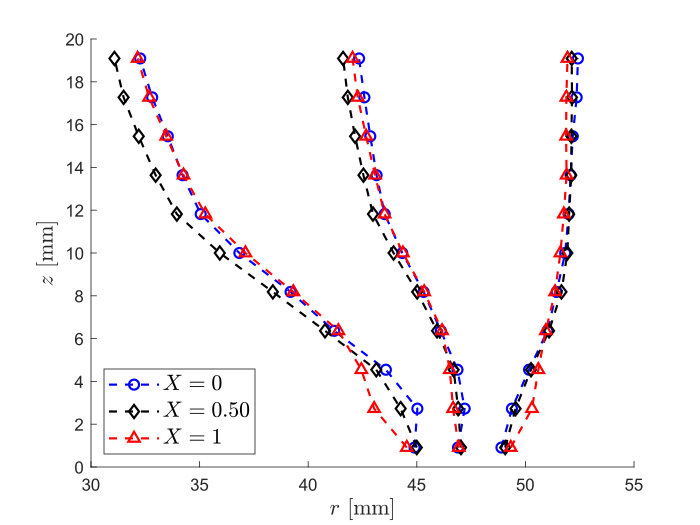}}; 
            \node[anchor=south east] at (0.4,2.8) {$R_c$};
            \node[anchor=south east] at (3.2,2.8) {$R_c+W$};
            \node[anchor=south east] at (-2,2.8) {$R_c-W$};
            \draw[dashed] (-2.9,1.78) -- (2.9,1.78);
            \node[anchor=south east] at (-0.2,1.7) {$z_\mathrm{top}$};
            \node[anchor=south east] at (3.2,-2.2) {(b)};
        \end{tikzpicture}
    \end{minipage}\hfill
                \vspace{30pt}
    \begin{minipage}{0.5\textwidth}
        \begin{picture}(100,170)
       \put(0,0){\includegraphics[width=1.0\textwidth]{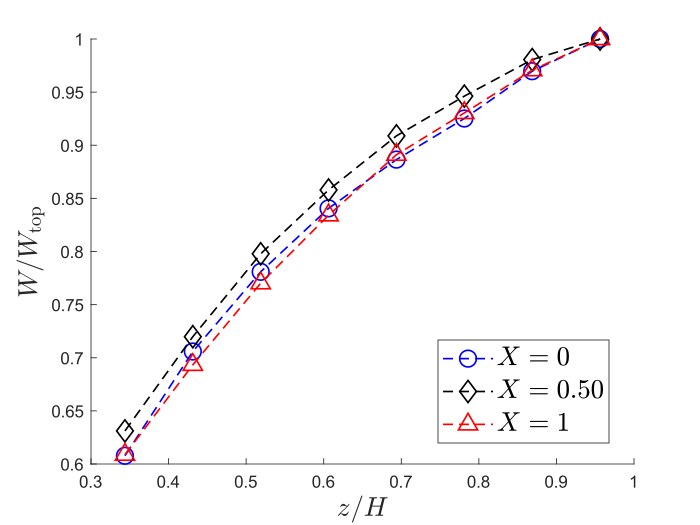}}
            \put(40,145){(c)}
        \end{picture}
    \end{minipage}\hfill
    \begin{minipage}{0.5\textwidth}
        \begin{picture}(100,170)
        \put(0,0){\includegraphics[width=1.0\textwidth]{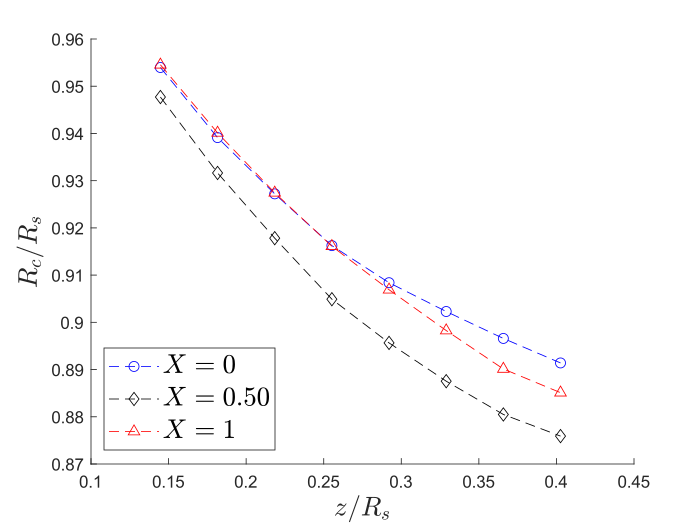}}
            \put(185,145){(d)}
        \end{picture}
    \end{minipage}\hfill
                    \vspace{30pt}
    \begin{minipage}{0.5\textwidth}
        \begin{picture}(100,170)
       \put(0,0){\includegraphics[width=1.0\textwidth]{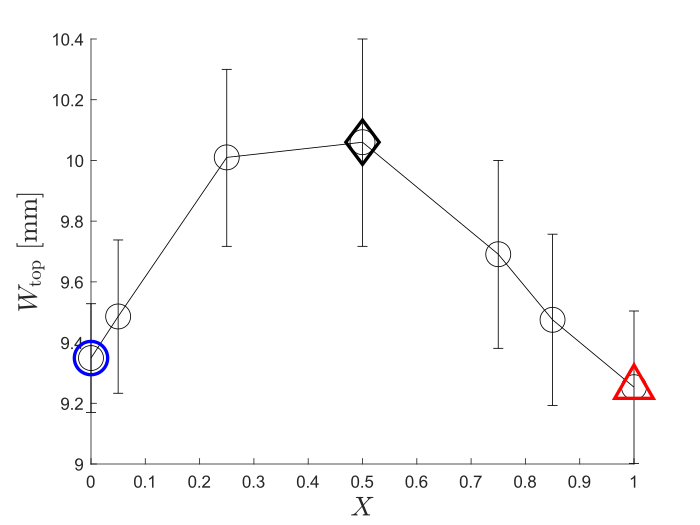}}
            \put(40,30){(e)}
        \end{picture}
    \end{minipage}\hfill
    \begin{minipage}{0.5\textwidth}
        \begin{picture}(100,170)
       \put(0,0){\includegraphics[width=1.0\textwidth]{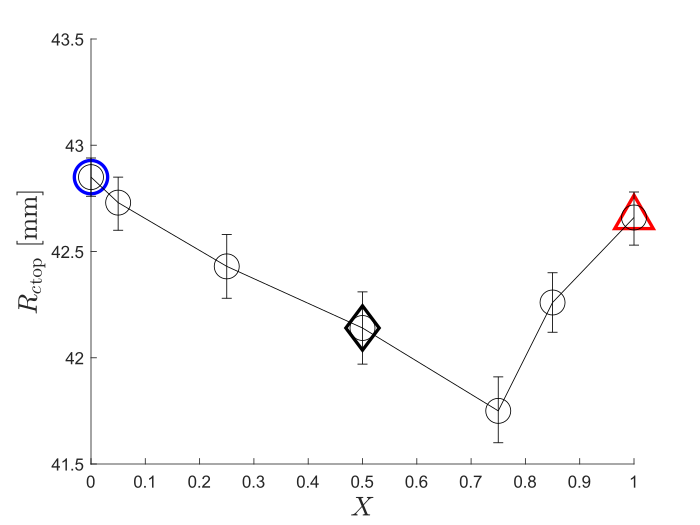}}
            \put(40,30){(f)}
        \end{picture}
    \end{minipage}\hfill
    \caption{(a) Contour plot of strain rate $\dot\gamma(r,z)$ as a function of the radial and vertical position for the sample with $X = 0.50$. The lines overlaid on the figure indicate the center $R_c$ and the edges $R_c+W$ and $R_c-W$ of the shear band as obtained from the fit function in \autoref{eq:SBSfit}. (b) Shear band center $R_c$ and edges $R_c+W$ and $R_C-W$ for different fractions of hard particles $X$, (c) scaled shear band width $W/W_\mathrm{top}$ as a function of scaled height $z/H$ and (d) relative position of shear band $R_c/R_s$ as a function of scaled height $z/R_s$ for samples with $X = 0, 0.50$ and $1$. (e) Shear band width $W$ as a function of the fraction of hard particles $X$ and (f) shear band position $R_c$ as a function of the fraction of hard particles $X$ at $z = z_\mathrm{top}$. The error bars in the graph illustrate the fitting error, which was determined by fitting the numerical data of $\omega/\omega_o$ vs. $r$ to the equation presented in \autoref{eq:SBSfit}. The blue {\color{blue}$\bigcirc$}, black $\diamondsuit$ and red {\color{red}$\triangle$} markers highlight the shear band width and position corresponding to $X = 0, 0.50$ and $1$, respectively.}\label{fig:shearbandwidth}
\end{figure*}

An important objective of this present study is to characterize the shear zone profiles of granular materials for ensembles of soft slippery and hard frictional materials. \autoref{fig:shearbandwidth}(b) shows the shear band profiles for samples with $X = 0, 0.50$ and $1$. As observed from the figure, the width of shear band corresponding to a mixture sample $X = 0.50$ is wider than that of the pure mixtures, corresponding to $X = 0$ and $1$. Also, the position of the shear band corresponding to a mixture sample $X = 0.50$ is shifted closer to the center than the pure mixtures. 
Note that the profile for the scaled width $W/W_\mathrm{top}$ as a function of $z/H$ shown in \autoref{fig:shearbandwidth}(c) does not follow a quadrant shape, as described by the analysis of Ries et al. \cite{ries2007shear}, for any of the samples. The reason behind this deviation needs further exploration. However, a possible reason for this deviation can be due to the low $W/d$ and $z/d$ ratios, where $d$ is the mean particle diameter. The normalized position of the shear band for the different mixtures as a function of bulk height is shown in \autoref{fig:shearbandwidth}(d), and the trends agree with that shown by Unger et al. \cite{unger2004shear}. From both \autoref{fig:shearbandwidth}(c) and (d), we observe a marked deviation in the trend for $X=0.50$ compared to the pure samples with $X = 0$ and $1$. Next, we analyze the width and position of the shear band for a horizontal slice through the axis of the shear cell at $z = z_\mathrm{top} = 16$ mm, as represented by the dashed line in \autoref{fig:shearbandwidth}(b), for all samples of different fraction of hard particles $X$ and show them in \autoref{fig:shearbandwidth}(e) and (f), respectively. Here, we find that the shear band gets wider as well as moves inwards in a mixture of soft and hard spheres up to a fraction $X = 0.50$, which corresponds to an intermediate mixture of $50$\% soft - $50$\% hard particles. With a further increase of $X$, the shear band narrows and moves outwards in the shear cell. The non-monotonic trend of the width of the shear band with increasing $X$ was also observed by Wang et al. \cite{wang2022characterization}.

From the above observations, we conclude that the ratio of hard spheres in a mixture of hard frictional and soft slippery particles significantly affects the flow profiles at the bulk scale. For up to intermediate ratios of hard particles, the shear band gets wider by up to $22\%$, and also the center shifts inwards by up to $2\%$, compared to the values for pure soft particles. Both of these features are influenced by the effect of inter-particle friction and stiffness, which are dominant for frictional hard particles \cite{luding2007effect}.

Wang et al. \cite{wang2022characterization} observed experimentally a notable difference in the shear band properties between samples made of pure materials and mixtures. It was observed that the shear zones of the mixtures are wider than those of the pure materials. Unlike the experimental observations, which show a significant widening of the shear zone for mixtures, our simulations show a rather small effect of widening the shear zone. However, the distinct non-monotonic trend of the shear band width as a function of the mixing ratio is the feature observed in experiments, which is also captured in our simulations. In addition, we observe the shift in shear band position for the mixtures, which were not reported in the experimental work. It is important to note that the capillary bridges resulting from small amounts of liquid between hydrogel particles potentially have an impact on experiments, despite not being taken into consideration in this numerical study. For larger quantities of liquid in the mixtures, beyond the capillary regime, the material would exhibit a more dispersed shear behavior.
\section{Simplified cases to explore the role of inter-particle friction and elastic modulus}

\begin{table*}
\small
\centering
  \caption{\ Simplified cases to explore the effects of friction and elastic modulus on shear band characteristics separately}
  \label{tab:CaseStudies1}
  \begin{tabular*}{0.5\textwidth}{@{\extracolsep{\fill}}lll}
    \hline
    Cases & Elastic modulus ($E$) & Friction coefficient ($\mu$)\\
    \hline
    $E_\mathrm{soft}$ $\mu_\mathrm{real}$ & $E = 50$ kPa for both species & $\mu_\mathrm{soft}=0.005$ $\vert$ $\mu_\mathrm{hard}=0.20$ \\
    $E_\mathrm{hard}$ $\mu_\mathrm{real}$ & $E = 5000$ kPa for both species & $\mu_\mathrm{soft}=0.005$ $\vert$ $\mu_\mathrm{hard}=0.20$ \\ 
    $E_\mathrm{real}$ $\mu_\mathrm{soft}$ & $E_\mathrm{soft} = 50$ kPa $\vert$ $E_\mathrm{hard} = 5000$ kPa & $\mu = 0.005$ for both species \\ 
    $E_\mathrm{real}$ $\mu_\mathrm{hard}$ & $E_\mathrm{soft} = 50$ kPa $\vert$ $E_\mathrm{hard} = 5000$ kPa & $\mu = 0.20$ for both species \\
    \hline
  \end{tabular*}
\end{table*}

Following the findings of this study and the qualitative agreement of our results with the observations noted by Wang et al. \cite{wang2022characterization}, it becomes evident that mixtures of soft, slippery and hard frictional grains demonstrate shear bands of variable width for increasing mixing ratio, which evolves non-monotonically with increasing values of the ratio. Since the materials of interest have two independent material parameters differing distinctly, namely friction and elastic modulus, it is interesting to look into their individual effects on the shear band characteristics. Hence, to explore the role of inter-particle friction and elastic modulus on the flow profile of the granular mixture systematically and independently of each other, we vary one parameter at a time in additional discrete element simulations of a parametric nature, while keeping the other parameter constant. Therefore, we conducted four case studies, as outlined in \autoref{tab:CaseStudies1}. In these studies, our primary focus is on analyzing the shear band characteristics at a specific height, denoted as $z = z_\mathrm{top}$ in \autoref{fig:shearbandwidth}(b) as a function of mixing ratio $X$. We then proceed to compare the obtained results with those from the real simulations.

\subsection{Shear band characteristics of granular mixtures}

Luding \cite{luding2007effect} examined the effects of friction on shear bands via simulations of a split-bottom shear cell. He showed that the presence of inter-particle friction does not affect the qualitative behavior of the shear band. However, the shear band moves inwards and gets narrower in the presence of friction. This is justified by the fact that with increasing friction, there is less relative displacement between the particles, particles tend to move more as a rigid block, and hence, the shear band width decreases. Here, we vary the inter-particle friction of the mixture by varying the fraction of hard particles $X$, keeping the elastic modulus constant in cases $E_\mathrm{soft}$ $\mu_\mathrm{real}$ and $E_\mathrm{hard}$ $\mu_\mathrm{real}$ as implied by the red and blue curves, respectively, in \autoref{fig:simplifiedCases}. For both the cases, we observe that the shear band width decreases with increasing $X$ in \autoref{fig:simplifiedCases}(a). Also, the shear band position moves inwards with increasing $X$ as shown in \autoref{fig:simplifiedCases}(b). Our results agree with the findings of Luding on the effects of friction on shear band characteristics.

On the contrary, varying the elastic modulus and considering same friction coefficient for the two materials leads to a wider shear band formation with increasing fraction of hard particles $X$, as demonstrated by cases $E_\mathrm{real}$ $\mu_\mathrm{soft}$ and $E_\mathrm{real}$ $\mu_\mathrm{hard}$. This is implied by the green and orange curves in \autoref{fig:simplifiedCases}(c). However, for the same friction between the two species, varying only the elastic modulus of the species has only a small effect on the position of the shear band as a function of $X$ as shown by the green and orange trends in \autoref{fig:simplifiedCases}(d).

\begin{figure*}
    \begin{minipage}{0.5\textwidth}
        \begin{picture}(100,170)
   \put(0,0){\includegraphics[width=1.0\textwidth]{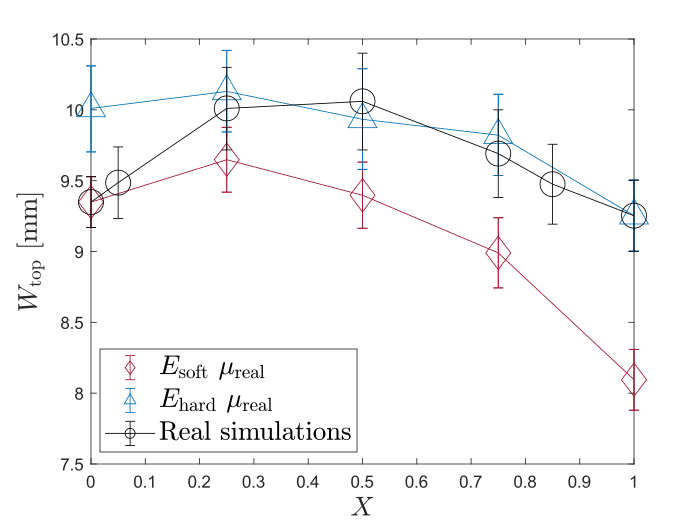}}
            \put(220,30){(a)}
        \end{picture}
    \end{minipage}\hfill
    \begin{minipage}{0.5\textwidth}
        \begin{picture}(100,170)
    \put(0,0){\includegraphics[width=1.0\textwidth]{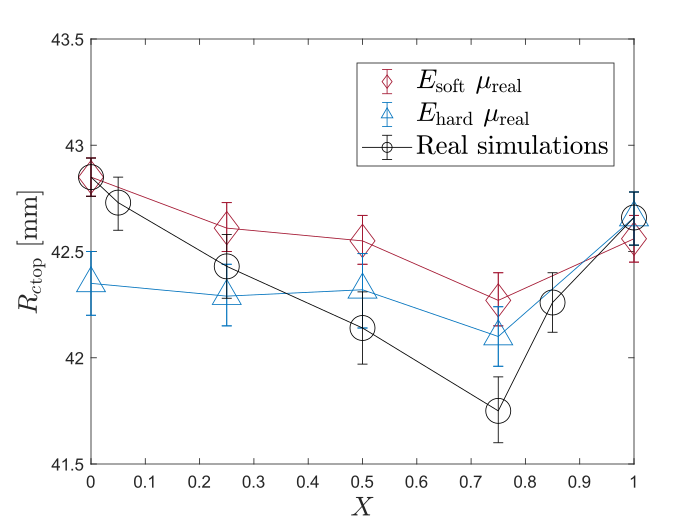}}
            \put(40,30){(b)}
        \end{picture}
    \end{minipage}

    \vspace{25pt}
    \begin{minipage}{0.5\textwidth}
        \begin{picture}(100,170)
     \put(0,0){\includegraphics[width=1.0\textwidth]{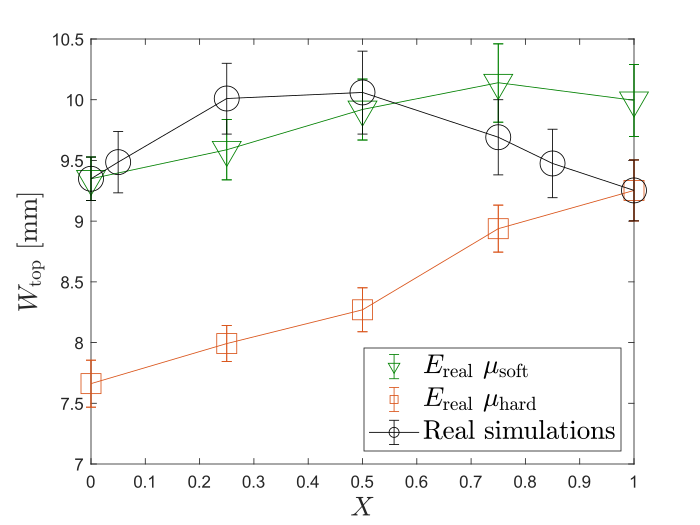}}
            \put(40,30){(c)}
        \end{picture}
    \end{minipage}\hfill
    \begin{minipage}{0.5\textwidth}
        \begin{picture}(100,170)
      \put(0,0){\includegraphics[width=1.0\textwidth]{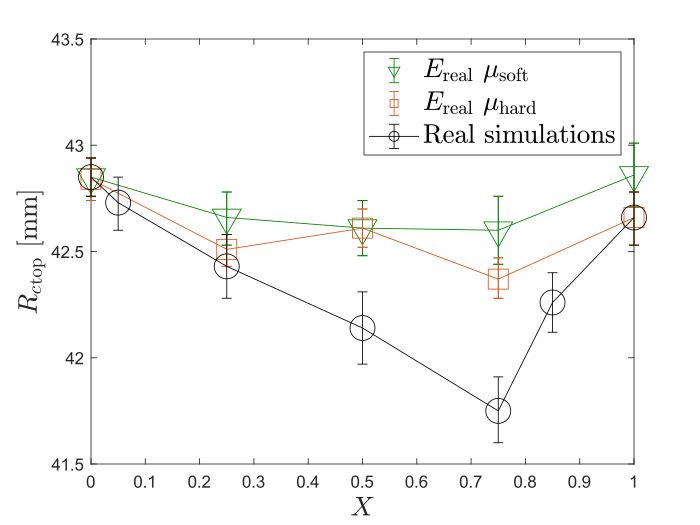}}
            \put(40,30){(d)}
        \end{picture}
    \end{minipage}\hfill
      \caption{Shear band (a) width $W$ and (b) position $R_c$ for fixed elastic modulus and real inter-particle friction, represented by cases $E_\mathrm{soft}$ $\mu_\mathrm{real}$ and $E_\mathrm{hard}$ $\mu_\mathrm{real}$, as compared with the real simulations, and shear band (c) width $W$ and (d) position $R_c$ for real elastic modulus and fixed inter-particle friction, represented by cases $E_\mathrm{real}$ $\mu_\mathrm{soft}$ and $E_\mathrm{real}$ $\mu_\mathrm{hard}$, as compared with the real simulations, where all the features are plotted as a function of fraction of hard particles $X$. The width and position of the shear band are obtained at $z_\mathrm{top} = 16$ mm. The error bars represent the fitting error, determined by fitting the numerical data for angular velocity to \autoref{eq:SBSfit}. }\label{fig:simplifiedCases} 
          \vspace{25pt}

\end{figure*}
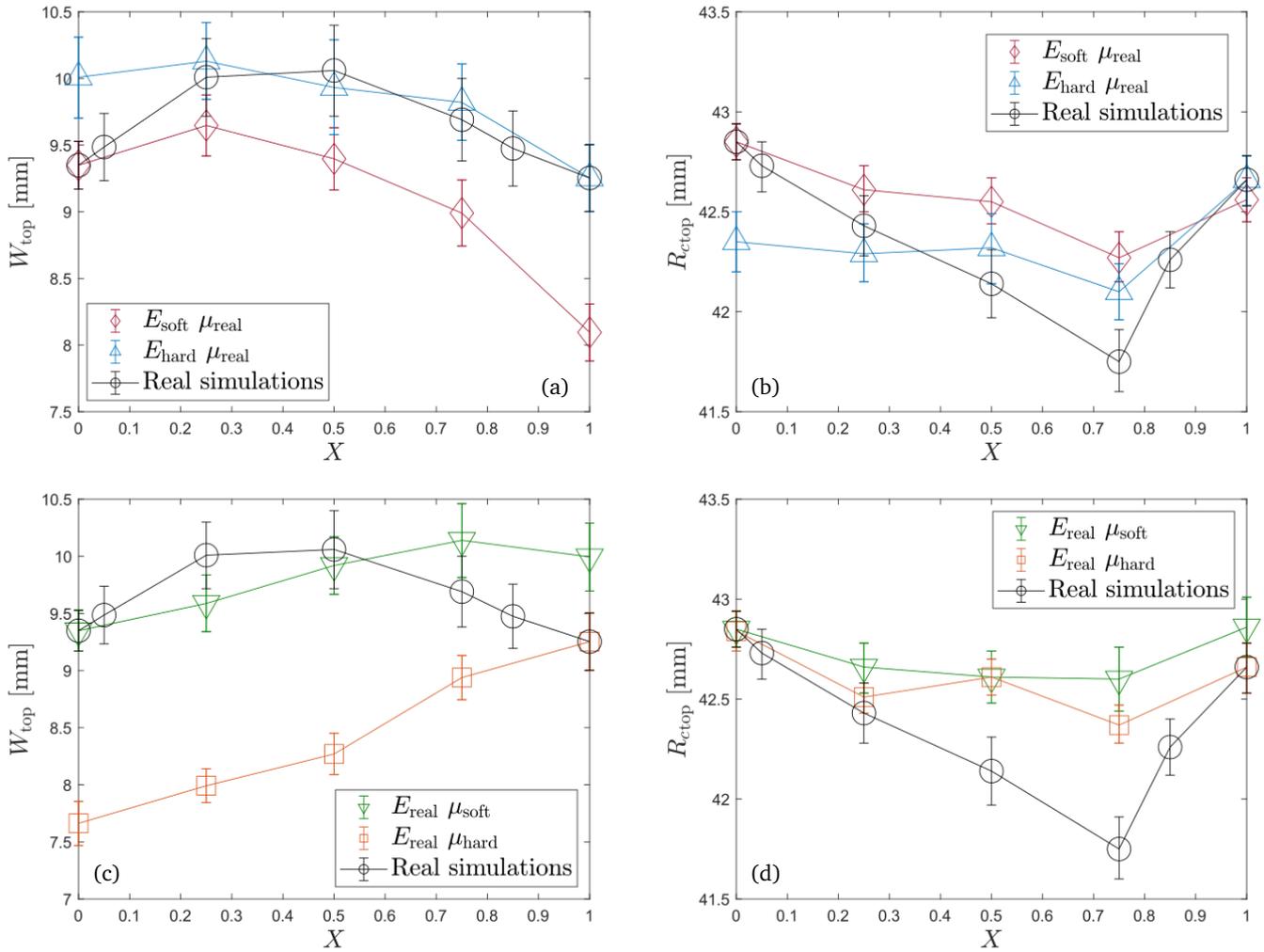

Overall, there is a dual effect of friction and elastic modulus that influences the shear band properties. While increasing friction tends to reduce the width of the shear band with increasing $X$, increasing elastic modulus plays the opposite role in increasing the width for increasing $X$. For smaller values of $X$, the friction of soft particles is smaller and hence is less dominant, i.e., it is easier for inter-particle contacts to slide. With increasing $X$, there are more contacts between hard frictional particles, and hence friction dominates over the effect of elastic modulus. 

\subsection{Macroscopic friction coefficient of granular mixtures}
The local shear stress $\tau(r,z)$ plotted against the local normal stress $P(r,z)$ is shown in \autoref{fig:muMixture}(a). Since the system is inhomogeneous in nature, for a given pressure, we observe a wide range of local strain rate values $\dot\gamma$, and we find that shear stress $\tau$ is higher for increasing strain rate $\dot\gamma$. However, for values above the critical strain rate, i.e., $\dot\gamma > \dot\gamma_c$, $\tau$ becomes almost independent of the local strain.
This means that $\tau/p$ is nearly constant for all data points with strain rate $\dot\gamma > \dot\gamma_c$. A linear trend is observed for the shear stress as a function of the normal stress, which is fitted well by:
\begin{equation}\label{eq:shearstress-fit}
    \tau = \mu P\,,
\end{equation}
where $\mu$ is the macroscopic friction coefficient obtained from linear fitting of the local shear stress and normal stress values.

We further analyze the macroscopic friction coefficient $\mu$ for different mixtures as a function of $X$ for the real simulations and the simplified cases and show the results in \autoref{fig:muMixture}(b). The macroscopic friction coefficient $\mu$ increases as a quadratic function with $X$ for the simplified cases of $E_\mathrm{soft}$ $\mu_\mathrm{real}$ and $E_\mathrm{hard}\mu_\mathrm{real}$, i.e., same elastic modulus for both materials and varying inter-particle friction coefficient. On the contrary, $\mu$ increases linearly with $X$ for cases $E_\mathrm{real}$ $\mu_\mathrm{soft}$ and $E_\mathrm{real}$ $\mu_\mathrm{hard}$, i.e. varying elastic modulus and same inter-particle friction coefficient for both materials. Furthermore, the macroscopic friction coefficient $\mu$ for the real simulations also increases as a quadratic function with $X$, which suggests that $\mu$ for the real simulations is strongly dominated by the varying inter-particle friction coefficient between the two species. The elastic modulus does not play a significant role in influencing the $\mu$ of the bulk behavior. Focusing on the functional form of the quadratic function $\mu$ vs $X$ for the real simulations, the black solid line represents the quadratic fitting equation $\mu = \mu_\mathrm{soft} + aX^2$, where $\mu_\mathrm{soft} \approx 0.14$ is the macroscopic friction coefficient corresponding to the pure soft particle mixtures and $a = 0.18$ here depends on the inter-particle friction coefficient values of pure soft and hard species.

\begin{figure*}
    \centering
    \begin{minipage}{0.5\textwidth}
        \begin{picture}(100,170)
    \put(0,0){\includegraphics[width=1.0\textwidth]{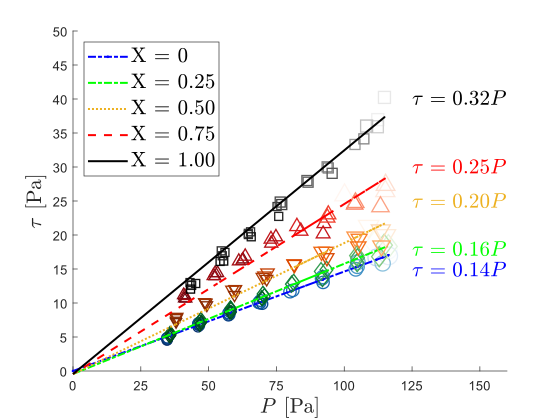}}
            \put(180,30){(a)}
            \end{picture}
            \end{minipage}\hfill
            \begin{minipage}{0.5\textwidth}
               \begin{picture}(100,170)
     \put(0,0){\includegraphics[width=1.0\textwidth]{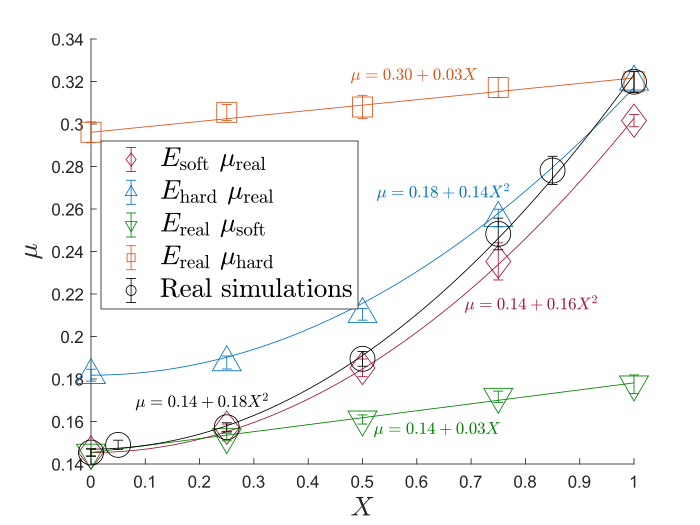}}
            \put(200,30){(b)}
        \end{picture}
    \end{minipage}\hfill
    \caption{(a) Shear stress $\tau$ plotted against normal stress $P$. The lines represent linear regression fits given by $\tau = \mu P$. (b) Macroscopic friction coefficient $\mu$ as a function of the fraction of hard particles $X$ for the real simulations and the simplified cases. The solid lines indicate linear and quadratic regression fits for the different cases. The error bars in the graph illustrate the fitting error, which was determined by fitting the numerical data of $\mu$ Vs $P$ to the equation presented in \autoref{eq:shearstress-fit}.}\label{fig:muMixture}
\end{figure*}

\section{Conclusions}
This work investigated the shear zone characteristics in binary mixtures of soft low frictional and hard frictional particles, via discrete element simulations of mixtures inside a split-bottom shear cell. 
The ratio of the volume of hard particles to the soft particles was varied parametrically to observe its effects on measures of interest, including the shear band width $W$, shear band position $R_c$ and the macroscopic friction coefficient $\mu$.

By comparing the width and position of the shear zone for samples of pure soft low frictional and for samples of pure hard frictional particles, it becomes evident that both systems show somewhat similar shear band features. However, mixtures of the two materials exhibit broader shear bands and a shifting of the shear band position towards the center of the shear cell. Both these effects demonstrate non-monotonicity for increasing fraction of hard particles. Although the changes in shear band characteristics for different mixtures are small, the non-monotonic trend of the shear band features as a function of mixing ratio is an interesting effect of the shear behavior of binary mixtures of this nature. We attribute the effects of shear band widening in these mixtures to an interplay of the role of inter-particle friction and particle stiffness.

Contrary to the shear band characteristics, the macroscopic friction coefficient shows a monotonically increasing trend as a quadratic function with an increasing fraction of hard particles in the mixtures. The macroscopic friction coefficient is strongly influenced by the inter-particle friction coefficient of the species present in the mixture, and the elastic modulus seems to play a less significant role.

This work captures critical features of shear band formation in soft-hard particle mixtures of low-high inter-particle friction, such as the non-monotonic evolution of the shear band width and position for increasing mixing ratios. However, the sensitivity of these features in our simulations is minor as compared to that observed in the experiments by Wang et al. \cite{wang2022characterization}. Some of the additional factors that might play a role in the experiments but are not captured in our simulations include the following: (i) our simulations only consider spherical particles and do not account for non-spherical shapes or the actual particle size distribution, (ii) the real stiffness of the mustard seeds is not taken into account; a sufficiently stiffer material is chosen for the mustard seeds instead, and (iii) capillary bridges due to small amounts of liquid between hydrogel particles might play a role in experiments which are not considered here. The material would exhibit a more dispersed shear behavior for larger quantities of water in the mixtures beyond the pendular regime. Even with these discrepancies, our work captures the shear band behavior observed experimentally and the evolution of its characteristics for varying mixing ratios of soft and hard particles.

\section*{Author Contributions}
SR and VA designed the research and planned the parametric DEM simulations which APS carried out. SR and VA wrote the first draft of the manuscript, and TP reviewed and wrote with SR and VA the second draft. All authors reviewed the results and approved the final version of the manuscript. SR secured the research funding.

\section*{Conflicts of interest}
There are no conflicts to declare.

\section*{Acknowledgements}
We acknowledge Stefan Luding for providing the code for micro-macro data transition and carefully reviewing this work. We thank Joshua Dijksman for stimulating discussion on this work. Financial support through the Emerging Talents Initiative (ETI Project No. ETI-Antrag\_2022-1\_Tech\_13\_Roy) and the DFG grant for Granular Weissenberg Effect (Project No. PO472/40-1) are acknowledged by SR and TP, respectively.



\balance


\bibliography{rsc,references} 
\bibliographystyle{rsc} 

\end{document}